\documentclass[3p,times]{elsarticle}

\usepackage{ecrc}

\volume{00}

\firstpage{1}

\journalname{Nuclear Physics A}

\runauth{M.~Bluhm et al.}

\jid{nupha}

\usepackage{amssymb}

\usepackage[figuresright]{rotating}

\begin{document}

\begin{frontmatter}



\dochead{}

\title{Radiative energy loss of relativistic charged particles in absorptive media}


\author{M.~Bluhm$^{1,2}$, P.~B.~Gossiaux$^1$, J.~Aichelin$^1$}

\address{$^1$ SUBATECH, UMR 6457, Universit\'{e} de Nantes,
Ecole des Mines de Nantes, IN2P3/CNRS. 4 rue Alfred Kastler,
F-44307 Nantes cedex 3, France}
\address{$^2$ CERN, Physics Department, Theory Division, 
CH-1211 Geneva 23, Switzerland}

\begin{abstract}
We determine the energy loss spectrum per time-interval of a relativistic charge 
traversing a dispersive medium. Polarization and absorption effects in the medium 
are modelled via a complex index of refraction. We find that the spectrum amplitude 
becomes exponentially damped due to absorption mechanisms. Taking explicitly the 
effect of multiple scatterings on the charge trajectory into account, we confirm 
results obtained in a previous work. 
\end{abstract}


\end{frontmatter}


\section{Introduction}
\label{sec:1}

The matter formed in high-energy heavy-ion collisions is an opaque QCD 
medium~\cite{Adcox02,Adler02,Aamodt10,Chatrchyan12,Steinberg11}, in which relativistic 
partons seem to suffer from a substantial medium-induced energy 
loss. Based on perturbative QCD calculations, this 
loss is commonly understood as dominated by the radiation of gluons off the energetic 
parton, cf.~\cite{Peigne09,Majumder11} for recent reviews. 
The properties of a medium can, however, significantly influence the probability 
for the formation of radiation and, in consequence, the amount of radiative energy 
loss~\cite{Ter-Mikaelian,Feinberg56,Klein99}. 
In~\cite{Bluhm11}, the impact of absorption and polarization effects in an infinite 
dispersive medium on the energy lost by a relativistic point-charge per unit length 
(or time) was investigated in linear response theory. These considerations 
are strictly valid only for electric charges in electro-magnetic plasmas, but 
can be understood as an Abelian approximation for the dynamics of a colour charge in 
a strongly interacting medium. The aim of the present work is to provide some more 
explanatory details for the results derived in~\cite{Bluhm11}. Thereby, we follow closely 
the reasoning of Landau and Pomeranchuk in their pioneering work~\cite{Landau53}, in 
particular, when including the effect of multiple scatterings the charge encounters 
during the radiation formation process. Throughout this work, natural units are used, 
i.e.~$\hbar=c=1$. 

\section{Derivation of the differential mechanical work}
\label{sec:2}

In line with the study in~\cite{Thoma91}, we determine the energy lost by a charge 
traversing an absorptive and polarizable medium from the negative of the work that has to be 
performed for moving against the electric field the charge induces in the medium. 
The work $W$ is conveniently evaluated in the mixed spatial coordinate $\vec{r}$ and frequency 
$\omega$ representation via 
\begin{equation}
\label{equ:Wmech2}
 W = 2 \,{\rm Re} \left(\int d^3 \vec{r}\,' \int_0^\infty d\omega \, 
 \vec{E}(\vec{r}\,',\omega) \, \vec{j}^*(\vec{r}\,',\omega) \right) .
\end{equation}
Here, $\vec{j}(\vec{r}\,',\omega)$ is the Fourier-transform of the real and time-dependent 
classical current $\vec{j}(\vec{r}\,',t)=q\vec{v}(t)\delta^{(3)}(\vec{r}\,'-\vec{r}(t))$ 
for a point-charge $q$ produced in the remote past and moving with velocity 
$\vec{v}(t)$, while $\vec{E}(\vec{r}\,',\omega)$ 
is the corresponding total electric field inside the medium as determined from Maxwell's 
equations for linear dispersive media. 

Considering in the following an isotropic and homogeneous medium, cf.~\cite{Ichimaru}, the 
constitutive expression for the electric field induced by the above current reads in 
Fourier-space 
\begin{equation}
\label{equ:DefinEfield}
 \frac{1}{\omega\mu(\omega)}\left[\vec{k}(\vec{k}\,\vec{E}_{\vec{k}}(\omega))
 -k^2\vec{E}_{\vec{k}}(\omega)\right]+\omega\epsilon(\omega)\vec{E}_{\vec{k}}(\omega) 
 = -\frac{iq}{(2\pi)^2}\int_{-\infty}^\infty dt' \vec{v}(t') 
 e^{i\omega t' -i\vec{k}\vec{r}(t')} \,.
\end{equation}
For simplicity, the permittivity $\epsilon(\omega)$ and permeability $\mu(\omega)$ of the medium 
are assumed to exhibit no momentum $\vec{k}$-dependence. Decomposing $\vec{E}_{\vec{k}}(\omega)$ 
as well as $\vec{v}(t)$ into components parallel and orthogonal to $\vec{k}$, 
Eq.~(\ref{equ:DefinEfield}) results in 
\begin{equation}
\label{equ:totalE}
 \vec{E}_{\vec{k}}(\omega) = 
 \frac{iq}{(2\pi)^2} 
 \int dt' \frac{e^{i\omega t' -i\vec{k}\vec{r}(t')}}{\omega\epsilon(\omega)} 
 \left\{\frac{(\vec{k}\,\vec{v}(t'))\vec{k}}{(\omega^2 n(\omega)^2-k^2)}-
 \frac{\vec{v}(t')\omega^2 n(\omega)^2}{(\omega^2 n(\omega)^2-
 k^2)} \right\} 
\end{equation}
for the total electric field, where $n(\omega)^2=\epsilon(\omega)\mu(\omega)$ 
denotes the $\omega$-dependent complex medium index of refraction squared. 

The negative work in differential form for positive $\omega$ following from 
Eq.~(\ref{equ:totalE}) together with $\vec{j}^*(\vec{r}\,',\omega)$ reads then 
\begin{equation}
\label{equ:Wmech3}
 - \frac{dW}{d\omega} = 
 {\rm Re} \,\Bigg(\frac{iq^2}{8\pi^4}\int dt \int dt' \int d^3\vec{k} \, 
 \frac{e^{-i\omega(t-t')+i\vec{k}\Delta \vec{r}}}{\omega\epsilon(\omega)} 
 \left\{\frac{(\vec{k}\,\vec{v}(t))(\vec{k}\,\vec{v}(t'))}{(k^2-
 \omega^2 n(\omega)^2)}-
 \frac{\vec{v}(t)\,\vec{v}(t') \omega^2 n(\omega)^2}{(k^2-\omega^2 
 n(\omega)^2)}\right\}
  \Bigg) \,,
\end{equation}
where $\Delta\vec{r}=\vec{r}(t)-\vec{r}(t')$. The three-momentum integral in 
Eq.~(\ref{equ:Wmech3}) can be evaluated analytically. In order to show this, we 
concentrate for the moment on the integral related to the first term in the parentheses 
of Eq.~(\ref{equ:Wmech3}) and may rewrite, cf.~\cite{Landau53}, 
\begin{equation}
\label{equ:momint1}
 \int d^3\vec{k} \frac{(\vec{k}\,\vec{v}(t))(\vec{k}\,\vec{v}(t'))}{(k^2-
 \omega^2 n(\omega)^2)} e^{i\vec{k}\Delta \vec{r}} = 
 -\frac{2\pi}{i} (\nabla_{\Delta \vec{r}}\,\vec{v}(t))
 (\nabla_{\Delta \vec{r}}\,\vec{v}(t')) \left(
 \frac{1}{\Delta r} \int_{-\infty}^\infty dk 
 \frac{k e^{ik\Delta r}}{(k^2-\omega^2 n(\omega)^2)} 
 \right) \,.
\end{equation}
The remaining $k$-integral is calculated by contour integration, 
where the contour has to be closed in the upper half complex momentum-plane because 
$\Delta r$ is positive semidefinit. Depending, in general, on the sign of the non-vanishing imaginary 
part of $n(\omega)=n_r(\omega)+i\,n_i(\omega)$, only one of the two simple poles in 
$k_{(0)}=sgn(n_i)\omega n(\omega)$ will be located within the closed contour. 
This results in 
\begin{equation}
\label{equ:momint2}
 \int_{-\infty}^\infty dk 
 \frac{k e^{ik\Delta r}}{(k^2-\omega^2 n(\omega)^2)} = 
 i \pi \exp[i\,sgn(n_i) \omega n_r(\omega) \Delta r]\cdot \exp[-\omega |n_i(\omega)| \Delta r] \,.
\end{equation}

The $\vec{k}$-integral related to the second term in the parentheses of 
Eq.~(\ref{equ:Wmech3}) can be evaluated in a similar fashion. This leads to our main 
result for the negative differential work for positive $\omega$ reading~\cite{Bluhm11} 
\begin{equation}
\label{equ:Wmech4}
 - \frac{dW}{d\omega} = 
 - {\rm Re} \,\Bigg(\frac{iq^2}{4\pi^2\omega\epsilon(\omega)}
 \int dt \int dt' e^{-i\omega(t-t')} 
 \left\{
 \omega^2 n(\omega)^2\vec{v}(t)\,\vec{v}(t')+
 (\nabla_{\Delta \vec{r}}\,\vec{v}(t))(\nabla_{\Delta \vec{r}}\,\vec{v}(t')) 
 \right\}
 \frac{e^{i\,sgn(n_i) \omega n_r(\omega) \Delta r - \omega |n_i(\omega)| \Delta r}}{\Delta r}
 \Bigg) \,.
\end{equation}
Irrespective of the sign of $n_i(\omega)$, the amplitude in 
Eq.~(\ref{equ:Wmech4}) is exponentially damped which implies a potential reduction 
of the negative differential work as compared to the case with a real valued medium index of 
refraction. In the limit $n_i(\omega)=0$ and $n_r(\omega)=1$, Eq.~(\ref{equ:Wmech4}) 
renders into the expression for the radiation intensity discussed in~\cite{Landau53} and, 
thus, represents a radiative energy loss spectrum. In an absorptive medium, 
$n_i(\omega)>0$, and in case $n_r(\omega)>0$ one can show that, considering for simplicity 
$\mu(\omega)=1$, Eq.~(\ref{equ:Wmech4}) also describes a loss of energy, which is the 
physical situation assumed in our study. 

As $\Delta\vec{r}\to -\Delta\vec{r}$ under variable interchange between $t$ and $t'$, 
only the part of the integrand in Eq.~(\ref{equ:Wmech4}) which is 
symmetric in $t$ and $t'$ gives rise to a non-vanishing contribution. Then, the 
time integrations in Eq.~(\ref{equ:Wmech4}) may be rearranged in such a way that, 
together with an overall factor of $2$, only $t\geq t'$ has to be considered in the 
$t$-integral. Furthermore, in order to stay in line with the original work of Landau and 
Pomeranchuk~\cite{Landau53}, one may neglect those terms in Eq.~(\ref{equ:Wmech4}) 
which are related to derivatives of $1/\Delta r$. Correspondingly, Eq.~(\ref{equ:Wmech4}) 
reduces to 
\begin{equation}
\label{equ:Wmech5}
 - \frac{dW}{d\omega} = 
 - {\rm Re} \,\Bigg(\frac{iq^2\omega n(\omega)^2}{2\pi^2\epsilon(\omega)}
 \int_{-\infty}^\infty dt' \int_{t'}^\infty dt \cos[\omega(t-t')] 
 \left\{\vec{v}(t)\,\vec{v}(t')-
 \frac{(\vec{v}(t)\,\Delta\vec{r})(\vec{v}(t')\,\Delta\vec{r})}{(\Delta r)^2} 
 \right\}
 \frac{e^{i\,sgn(n_i) \omega n_r(\omega) \Delta r - \omega |n_i(\omega)| \Delta r}}{\Delta r}
 \Bigg) \,.
\end{equation}

Such an approximation might at first appear reasonable for large $\Delta r\gg 1$. It 
is worthwhile noting, however, that taking into account the terms that are omitted by 
such a procedure is, in principle, of equivalent importance. This can easily be seen, 
for instance, by realizing that in~\cite{Landau53} the correct limiting Bethe-Heitler 
radiation spectrum can only be obtained if one assumes a mean squared transverse 
momentum transfer that is twice as large as the one for a single 
Rutherford-scattering. This implies, as was already discussed in~\cite{Blankenbecler96}, 
that the terms omitted in~\cite{Landau53} give exactly the same contribution as those 
that were taken into account in~\cite{Landau53}. 

\section{Result for a specific charge trajectory}
\label{sec:3}

The important quantity entering Eqs.~(\ref{equ:Wmech4}) and~(\ref{equ:Wmech5}) is $\Delta\vec{r}$, 
which is determined for a given $\vec{v}(t)$ via 
$\Delta\vec{r}=\int_{t'}^t\vec{v}(\tau)d\tau$. In case of a time-independent velocity this 
results in $\Delta\vec{r}=\vec{v}\cdot(t-t')$, such that the differential work in 
Eq.~(\ref{equ:Wmech5}) exactly vanishes. For a time-dependent $\vec{v}(t)$, one may rewrite 
the time-integrals appearing in Eq.~(\ref{equ:Wmech5}) as 
\begin{eqnarray}
\nonumber
 \mathcal{J} & \equiv & \int_{-\infty}^\infty dt' \int_{t'}^\infty dt \cos[\omega(t-t')] 
 \left\{\vec{v}(t)\,\vec{v}(t')-
 \frac{(\vec{v}(t)\,\Delta\vec{r})(\vec{v}(t')\,\Delta\vec{r})}{(\Delta r)^2} 
 \right\}
 \frac{e^{i\,sgn(n_i) \omega n(\omega) \Delta r}}{\Delta r} \\
\label{equ:jint1}
 & = & \int_{-\infty}^\infty dt' \int_{0}^\infty d\bar{t} \cos[\omega\bar{t}\,] 
 \left\{\vec{v}(t')\,\vec{v}(t'+\bar{t}\,)-
 \frac{(\vec{v}(t')\,\Delta\vec{r})(\vec{v}(t'+\bar{t}\,)\,\Delta\vec{r})}{(\Delta r)^2} 
 \right\}
 \frac{e^{i\,sgn(n_i) \omega n(\omega) \Delta r}}{\Delta r}
\end{eqnarray}
by substituting $\bar{t}=t-t'$, where now $\Delta\vec{r}=\int_{0}^{\bar{t}}\vec{v}(t'+\tau)d\tau$. 

In order to quantify $\mathcal{J}$ in Eq.~(\ref{equ:jint1}), one may assume that any given initial 
velocity $\vec{v}(t')$ is changed by multiple scatterings in the medium according to 
$\vec{v}(t'+\tau)=\vec{v}(t')\cos\theta(\tau)+|\vec{v}(t')|\,\vec{e}_\perp\sin\theta(\tau)$~\cite{Landau53}. 
Here, $\vec{e}_\perp$ is a unit vector perpendicular to $\vec{v}(t')$ and $\theta(\tau)$ 
is the angle relative to $\vec{v}(t')$ that the charge accumulates through these scatterings 
within the time-interval between $t'$ and $t'+\tau\geq t'$. For this specific form of 
$\vec{v}(t)$, one finds 
$\vec{v}(t')\,\vec{v}(t'+\bar{t}\,)=v(t')^2\cos\theta(\bar{t}\,)\simeq v(t')^2(1-\theta(\bar{t}\,)^2/2)$ 
in the approximation of small deflection angles within the time-duration $\bar{t}$ after $t'$. 
Likewise, one finds 
$\Delta\vec{r}\simeq \vec{v}(t')(\bar{t}-\mathcal{I}_2/2)+v(t')\,\vec{e}_\perp\mathcal{I}_1$ for 
small deflection angles, where $\mathcal{I}_2 = \int_0^{\bar{t}}\theta(\tau)^2 d\tau$ 
and $\mathcal{I}_1 = \int_0^{\bar{t}}\theta(\tau) d\tau$. 

Assuming random kicks from the medium constituents, stochastic averaging over the deflection angles yields, 
for instance, $\langle\mathcal{I}_2\rangle =\hat{q}\bar{t}^{\,2}/E^2$, where the quantity $\hat{q}$ is 
defined here as $\hat{q}=\langle\theta(\tau)^2\rangle E^2/(2\tau)$, i.e.~one half of the mean of the 
transverse momentum transfer squared per unit 
time\footnote{We note that in Ref.~\cite{Bluhm11} the same quantity $\hat{q}$ was incorrectly identified 
with the mean of the transverse momentum transfer squared per unit time. With the common definition 
of $\hat{q}=\langle\theta(\tau)^2\rangle E^2/\tau$, the expression for $-dW/d\omega$ found in~\cite{Bluhm11} 
for the specific charge trajectory considered here and in~\cite{Bluhm11} is modified to some extent, 
see~\cite{Bluhm12}.}, and $E$ is the energy of the charge. As a consequence, one obtains 
$\langle\Delta r\rangle\simeq v(t')\bar{t}\sqrt{1-\hat{q}\bar{t}/(3E^2)}$. Omitting terms of 
order $\mathcal{O}(\theta^4)$ and approximating the exact average of $\mathcal{J}$ by taking 
the mean of each individual term entering Eq.~(\ref{equ:jint1}), we find 
\begin{equation}
\label{equ:jint2}
 \langle\mathcal{J}\rangle\simeq -\frac{\hat{q}}{3E^2} \int_{-\infty}^\infty dt' v(t') 
 \int_{0}^\infty d\bar{t} \cos[\omega\bar{t}\,] 
 e^{i\,sgn(n_i) \omega n(\omega) \langle\Delta r\rangle} \,,
\end{equation}
where here $\langle\Delta r\rangle\simeq v(t')\bar{t}(1-\hat{q}\bar{t}/(6E^2))$. Inserting 
$\langle\mathcal{J}\rangle$ from Eq.~(\ref{equ:jint2}) into Eq.~(\ref{equ:Wmech5}) 
finally results in 
the energy loss spectrum per time-interval presented in~\cite{Bluhm11}. We 
note that the expression obtained in Eq.~(\ref{equ:jint2}) is a direct consequence of both 
the particular form of $\vec{v}(t)$ and the assumption of small deflection angles. The 
latter gives rise to the condition $\hat{q}\bar{t}/(3E^2)\ll 1$ which implies that, for 
a given $E$, $\bar{t}$ in the integral in Eq.~(\ref{equ:jint1}) may not be considered 
too large or, otherwise, $\langle\Delta r\rangle$ becomes unphysical. 
This condition is, however, naturally satisfied as the relevant 
contributions to the $\bar{t}$-integral in Eq.~(\ref{equ:jint1}) stem from the 
region, in which the oscillating functions vary only slowly, i.e.~for $\bar{t}$ 
smaller than or of the order of the formation time, cf.~\cite{Bluhm11,Bluhm12}. 

\section{Characterizing the absorptive medium by a complex index of refraction}
\label{sec:4}

Absorption and polarization effects in the dispersive medium can effectively be 
taken into account via a complex index of refraction~\cite{Ichimaru}. 
Assuming that the radiated quanta obey a dispersion relation that is influenced by 
the medium, a suitable ansatz for a momentum-independent $n(\omega)^2$ is 
\begin{equation}
\label{equ:index}
 n(\omega)^2=1-\frac{m^2}{\omega^2}+2i\frac{\Gamma}{\omega} \,.
\end{equation}
In this way, one attributes both an effective mass $m$~\cite{Kampfer00} and, as 
result of damping mechanisms, a finite width $\Gamma$ to the quanta. In general, 
$m$ and $\Gamma$ can be free parameters of a Lorentz-type spectral function for 
these excitations~\cite{Peshier0405}. 

A consequence of the special ansatz Eq.~(\ref{equ:index}) is that radiation cannot be emitted 
for $\omega<m$. This follows directly from the dispersion relation as determined 
by ${\rm Re}(\omega^2 n(\omega)^2-k^2)=0$. In addition, the latter implies also that 
the emitted quanta are time-like excitations with $\omega$, for which $n_i(\omega)$ 
does not vanish. In~\cite{Thoma91,Djordjevic03}, in contrast, the dielectric functions 
for a strongly interacting medium were determined from the leading-order 
hard-thermal-loop gluon self-energy. In this case, emitted gluons are also time-like 
quanta with $\omega$, however, for which $n_i(\omega)=0$. 

\section{Conclusions}
\label{sec:5}

In this work, we presented a more detailed derivation of the results obained 
in~\cite{Bluhm11} for the energy loss spectrum per time-interval of a relativistic charge 
traversing an infinite dispersive medium. Absorption mechanisms were found to result 
in an exponential damping of the spectrum amplitude. Our classical study, which takes 
as in~\cite{Landau53} 
multiple scatterings of the charge in the medium into account, is restricted 
to the regime of small $\omega\ll E$ and is strictly valid only for describing electro-magnetic 
phenomena. Missing the important non-Abelian effect of gluon rescatterings, our 
approach may, nevertheless, be viewed as an Abelian approximation for the dynamics 
of colour charges in the absorptive quark-gluon plasma. 

Our main result for the negative differential work given in Eq.~(\ref{equ:Wmech4}) was obtained 
by assuming that the charge was created in the remote past and that the complex index of 
refraction was $\vec{k}$-independent. Such simplifications are, however, 
unrealistic for the situation encountered in high-energy heavy-ion collisions. In particular, 
the interference between the medium induced radiation and the initial bremsstrahlung due to 
the production process of the charge within the medium may be of importance. 
Consequently, the considerations presented here will necessarily have to be 
improved before firm quantitative conclusions can be drawn. We leave these issues to be 
addressed in a forthcoming publication. 

\section*{Acknowledgements}

MB thanks M.~Nahrgang for fruitful discussions and the organizers of the Hard Probes 2012 
conference for the financial support. 






\begin{thebibliography}{00}


\bibitem{Adcox02} K.~Adcox et al. (PHENIX), Phys. Rev. Lett. {\bf 88} (2002) 022301. 
\bibitem{Adler02} C.~Adler et al. (STAR), Phys. Rev. Lett. {\bf 89} (2002) 202301. 
\bibitem{Aamodt10} K.~Aamodt et al. (ALICE), Phys. Lett. B {\bf 696} (2011) 30. 
\bibitem{Chatrchyan12} S.~Chatrchyan et al. (CMS), Eur. Phys. J. C {\bf 72} (2012) 1945. 
\bibitem{Steinberg11} P.~Steinberg et al. (ATLAS), J. Phys. G {\bf 38} (2011) 124004. 
\bibitem{Peigne09} S.~Peign\'{e} and A.~V.~Smilga, Phys. Usp. {\bf 52} (2009) 659. 
\bibitem{Majumder11} A.~Majumder and M.~van~Leeuwen, Prog. Part. Nucl. Phys. A {\bf 66} (2011) 41. 
\bibitem{Ter-Mikaelian} M.~L.~Ter-Mikaelian, Dokl. Akad. Nauk SSSR {\bf 94} 
(1954) 1033; {\it High-Energy Electromagnetic Processes in Condensed Media} 
(John Wiley \& Sons, New York, 1972). 
\bibitem{Feinberg56} E.~L.~Feinberg and I.~Ya.~Pomeranchuk, Suppl. Nuovo Cimento {\bf 3} (1956) 652. 
\bibitem{Klein99} S.~Klein, Rev. Mod. Phys. {\bf 71} (1999) 1501. 
\bibitem{Bluhm11} M.~Bluhm, P.~B.~Gossiaux, and J.~Aichelin, Phys. Rev. Lett. {\bf 107} (2011) 265004. 
\bibitem{Landau53} L.~D.~Landau and I.~Ya.~Pomeranchuk, Dokl. Akad. Nauk SSSR {\bf 92} (1953) 535; 
 ibid. {\bf 92} (1953) 735. 
\bibitem{Thoma91} M.~H.~Thoma and M.~Gyulassy, Nucl. Phys. B {\bf 351} (1991) 491. 
\bibitem{Ichimaru} S.~Ichimaru, {\it Basic Principles of Plasma Physics} (W.~A.~Benjamin, 
Reading, 1973). 
\bibitem{Blankenbecler96} R.~Blankenbecler and S.~D.~Drell, Phys. Rev. D {\bf 53} (1996) 6265. 
\bibitem{Bluhm12} M.~Bluhm, P.~B.~Gossiaux, T.~Gousset and J.~Aichelin, arXiv:1204.2469 [hep-ph]. 
\bibitem{Kampfer00} B.~K\"ampfer and O.~P.~Pavlenko, Phys. Lett. B {\bf 477} (2000) 171. 
\bibitem{Peshier0405} A.~Peshier, Phys. Rev. D {\bf 70} (2004) 034016; 
 J. Phys. G {\bf 31} (2005) S371. 
\bibitem{Djordjevic03} M.~Djordjevic and M.~Gyulassy, Phys. Rev. C {\bf 68} (2003) 034914; 
 Phys. Lett. B {\bf 560} (2003) 37. 

\end{thebibliography}



\end{document}